# Study for Performance of MobileNetV1 and MobileNetV2 Based on Breast Cancer

**Jiuqi Yan**[1, *]

Electrical and Electronic Engineering The University of Hong Kong, Room 504 Hong Kong Baptist Church, Hong Kong, 2859 2111

[*]Corresponding author's email: yjqhku@connect.hku.hk

**Abstract.** Artificial intelligence is constantly evolving and can provide effective help in all aspects of people's lives. The experiment is mainly to study the use of artificial intelligence in the field of medicine. The purpose of this experiment was to compare which of MobileNetV1 and MobileNetV2 models was better at detecting histopathological images of the breast downloaded at Kaggle. When the doctor looks at the pathological image, there may be errors that lead to errors in judgment, and the observation speed is slow. Rational use of artificial intelligence can effectively reduce the error of doctor diagnosis in breast cancer judgment and speed up doctor diagnosis. The dataset was downloaded from Kaggle and then normalized. The basic principle of the experiment is to let the neural network model learn the downloaded data set. Then find the pattern and be able to judge on your own whether breast tissue is cancer. In the dataset, benign tumor pictures and malignant tumor pictures have been classified, of which 198738 are benign tumor pictures and 78, 786 are malignant tumor pictures. After calling MobileNetV1 and MobileNetV2, the dataset is trained separately, the training accuracy and validation accuracy rate are obtained, and the image is drawn. It can be observed that MobileNetV1 has better validation accuracy and overfit during MobileNetV2 training. From the experimental results, it can be seen that in the case of processing this dataset, MobileNetV1 is much better than MobileNetV2.

**1. Introduction**

In research released in 2020 by the World Health Organization's International Agency for Research on Cancer (IARC), breast cancer has supplanted lung cancer as the most common cancer globally [1]. The majority (about 80%) of breast cancer cases are of the invasive ductal carcinoma (IDC) variety. Among women, breast cancer has the highest proportion of cancer patients, far exceeding other cancer types. Reducing mortality from breast cancer requires early detection and diagnosis. However, the treatment process of breast cancer is extremely complex and unpredictable. Typically, doctors make diagnosis based on their own experience, but the treatment plans for the same patient at different stages are different. This makes the diagnosis inefficient and sometimes prone to This can lead to misdiagnosis and thus worsening of the disease.

In recent years, with the rapid development of computer computing power and artificial intelligence, deep learning have been employed everywhere, especially in image processing, where great progress has been made [2, 3]. Deep learning techniques can be used to automatically extract features from images, avoiding the limitations of manual feature extraction in traditional machine learning and saving manpower. Nowadays, many scholars have applied deep learning techniques in breast cancer diagnosis, which has improved the accuracy of breast cancer diagnosis to some extent. Spanhol et al.

applied AlexNet network on BreaKHis dataset and obtained satisfactory result compared with traditional methods [4]; Zou Wenkai et al. adapted the Inception structure of GoogleNet and used the method of uniform training and independent testing for all magnifications, with patient level as the evaluation criterion. The accuracy rate was 87%-90% [5]. Tan et al. performed a preprocessing operation to convert mammograms into computer-recognized images for normal, benign tumor and cancer mammograms, and used CNN to detect and classify these images [6]. The experimental results showed that the classification accuracy could reach 82.71%. Zhang et al. developed different CNN models to classify 2D mammograms and 3D tomosynthetic images, and evaluated each classification based on biopsy and expert confirmation [7]. Each classifier was evaluated based on biopsy and expert confirmation. However, the number of data sets used in these schemes is too small, which may affect the accuracy of the results.

To solve the limitation mentioned above, this study hopes to design a system that can automatically diagnose breast cancer. There are several steps in breast cancer diagnosis, including palpation, mammography or ultrasound imaging and breast tissue biopsy. The most important step in diagnosing this situation with IDC is to perform a graded operation on the aggressiveness of the IDC. To do so, pathologists usually focus on the regions of the mount sample where IDC is present. The automation of the detection of the exact regions of IDC inside of a whole mount slide could help reduce costs and time of the test.

The goal of machine learning, as the name implies, is to enable computers to imitate the learning behaviour of the human brain structure, to integrate existing knowledge with new knowledge, and to improve their own performance and structure. In the experiments, three neural networks were used, namely MobileNet V1 and MobileNet V2, to study the effects of different network structures on the detection accuracy.

## 2. Methodology

*2.1. Data preparation*
The data was downloaded from Kaggle [8], the study have a total of 277, 524 image files. 198, 738 of the patches are negative to IDC, while 78, 786 patches contain IDC. Figure 1 provides 15 images from the dataset.

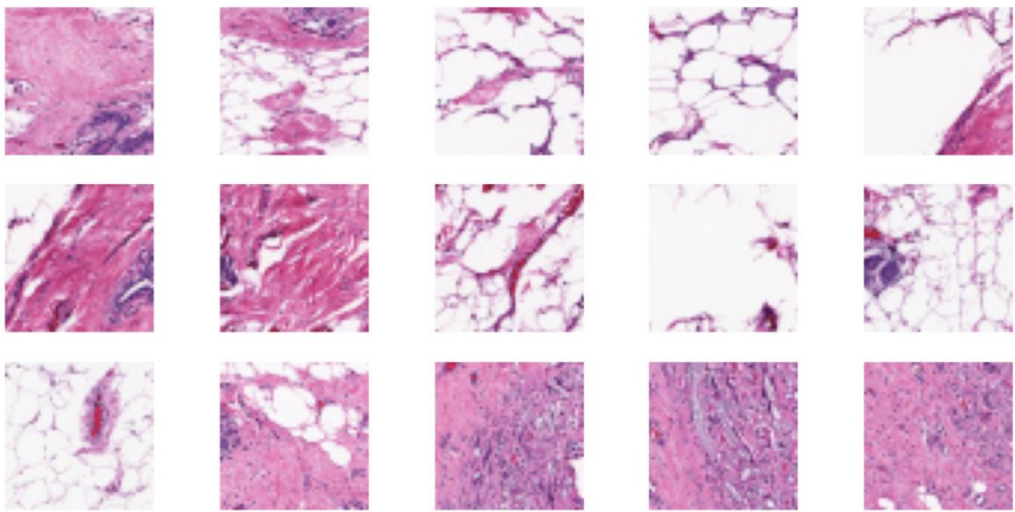

**Figure 1.** Sample part of the dataset.

*2.2. Introduction for CNN models*
Large-scale image processing can be accomplished with convolutional neural networks, a type of feed-forward neural network where artificial neurons can react to nearby units. A convolutional neural network consists of a convolutional layer and a pooling layer. Convolutional neural networks have a total of five-layer structures, including convolutional layer, activation layer, pooling layer, and fully connected FC layer.

Convolutional neural networks' core layer, which generates the majority of the network's computation, is the convolutional layer. First, the convolutional layer needs to implement the local perception function. As the human brain recognizes the function of pictures, prior to the upper level performing a thorough operation on the local area to obtain the global information, it all first senses each feature of the image locally. The computational parameters of the model are greatly reduced by the local perception feature. With the local perception feature, the computational parameters of the model are greatly reduced. But this alone still leaves a lot of parameters, so weight sharing is needed. Weight sharing makes ensuring that connections are distributed evenly across all points in the same tier.

The layer's activation function creates a nonlinear mapping of the convolution layer's output. Common activation functions include Sigmoid, Tanh and ReLU, etc.

Pooling: often referred to as downsampling or undersampling. Feature dimensionality reduction, data and parameter compression, overfitting reduction, and model fault tolerance improvement are its key applications. Max Pooling and Average Pooling are the two main ones. Finally, the output layer, the model inserts a high-quality feature picture learned into the fully connected layer following a number of prior convolutions, excitations, and pooling. The final output is produced by a softmax function, and the completely connected layer can be thought of as a straightforward multi-classification neural network (e.g. a BP neural network).

*2.3. Introduction for MobileNetV1 model and MobileNetV2 model*

Depthwise Separable Convolution used in MobileNetV1 is one of the most classical strategies for model compression, which is achieved by replacing 3×3 convolution across channels with 3×3 convolution in a single channel + 1×1 convolution across channels.

MobileNetV2 is based on the Depthwise Separable of MobileNetV1 and introduces the residual structure. And ReLU was found to have very serious information loss problem on Feature Map with small number of channels, which led to the introduction of Linear Bottlenecks and Inverted Residual.

MobileNetV1 is mainly used by deep separable convolution in the Inception model to reduce the computational effort of the first few layers. The technical route of MobileNetV1 focuses on the decomposition of the network into a depth-separable convolution (i.e. a depthwise convolution and a pointwise convolution) and the reduction of the number of channels (Thinner Models Although) and the resolution of the input (Reduced Representation) by two superparameters, width multiplier $\alpha$ and resolution multiplier $\beta$, respectively. "Separable" refers to the division into depthwise convolution and pointwise convolution of the conventional two-step filtering and combining in convolution. Each input channel is given a filter during depthwise convolution. Each depthwise convolution's output is subjected to a linear combination during the pointwise convolution.

MobileNetV2 fully maintains the structural simplicity of MobileNetV1, which means that the same accuracy can be achieved without adding additional special operations. In order to create the most straightforward network architecture feasible, MobileNetV2 is focused on investigating how neural networks function. It is based on two main areas of research: the use of optimization methods, such as genetic algorithms and reinforcement learning, to perform framework searches [9]; and the treatment of the "BottleNeck" Structure [10]. In addition to following the 3×3 depth-separable convolution, the two main innovations of MobileNetV2 are Line bottlenecks linear bottlenecks and Inverted residuals inverted residuals. The reason for introducing the linear bottleneck is that the "manifold of interest" region, which can be non-zero after ReLU, is also approximately linearly transformed. Second, after ReLU, some channel information will be lost. The reason for introducing inverted residuals: firstly, the bottleneck already contains the necessary information, so the shortcut directly connects the two bottlenecks. secondly, the information in the low-dimensional space is more important, so the scheme of first up-dimensioning and then extracting features and then down-dimensioning is used.

In this paper, MobileNetV1 and MobileNetV2 are used to train the dataset respectively. Since not all images have the same size, this experiment will need to make the sizes all equal so that the proposed network can work at its best. Before processing the data, the images are subjected to resize. This study then imports two neural network models and add a pooling layer and a fully connected layer to refine the network model.

*2.4. Implementation details*
This experiment was done on a laptop containing a Nvidia GTX1070. The program is done using the Tensorflow-GPU. The learning rate is 0.0001 and the epochs are 5. The optimizer is Adam, the evaluation metric is accuracy, and the loss function is cross-entropy loss function.

## 3. Result and discussion
In this work, the training effects of MobileNetV1 and MobileNetV2 on the same dataset and the accuracy of the test were compared.

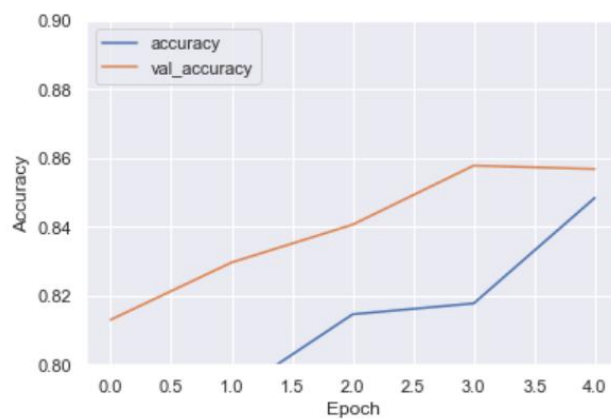

**Figure 2.** Experimental results of MobileNetV1.

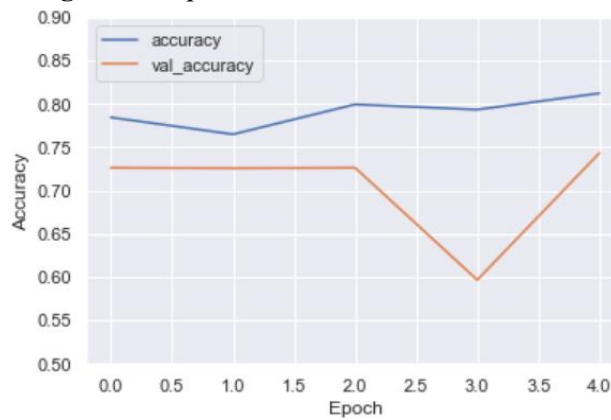

**Figure 3.** Experimental results of MobileNetV1.

**Table 1.** Experimental results.

|  | **Accuracy** | **Val_accuracy** |
| --- | --- | --- |
| MobileNetV1 | 0.849 | 0.857 |
| MobileNetV2 | 0.819 | 0.743 |

In Figure 2, it can be observed that the detection accuracy obtained from the training is gradually increasing and finally reaches about 85.6%. And in Figure 3, observing the curve of detection accuracy, it can be found that the detection accuracy obtained from the first three training rounds is nearly the same value, all about 72.6%, while it suddenly drops to about 59.6% in the fourth round, and then reaches 74.3% in the fifth round.

For the reason about the decrease of performance based on the MobileNetV2, it can be though that the structural parameters of MobileNetV2 are increased compared to MobileNetV1 and may not be appropriate for the processing of this dataset since it is prone to overfit. In addition, the appearance of residual block in MobileNetV2 may be beneficial for larger dataset instead of the current data used in the study. Therefore, the detection accuracy of MobileNetV2 is not as good as that of MobileNetV1, and the cut detection accuracy curve is very unstable.

## 4. Conclusion

This study was to compare which of MobileNetV1 and MobileNetV2 models was better at detecting histopathological images of the breast downloaded at Kaggle. For this experiment, only two neural network models, MobileNetV1 and MobileNetV2, were used for comparative analysis. Future study will add MobileNetV3 and ResNet for another comparative analysis, and it can change the dataset, such as using other breast cancer CT images, or using different disease images such as brain cancer images and lung cancer images.